\documentclass[twocolumn,showkeys,showpacs,nofootinbib]{revtex4}
\usepackage{amsfonts}
\usepackage{amsmath}
\usepackage{amsxtra}
\usepackage{graphicx}

\begin{document}


\title{Thermodynamics in dynamical spacetimes}


\author{Romualdo Tresguerres}
\email[]{romualdotresguerres@yahoo.es}
\affiliation{Instituto de F\'isica Fundamental\\
Consejo Superior de Investigaciones Cient\'ificas\\ Serrano 113
bis, 28006 Madrid, SPAIN}

\date{\today}

\begin{abstract}
We derive a general formulation of the laws of irreversible thermodynamics in the presence of electromagnetism and gravity. For the handling of macroscopic material media, we use as a guide the field equations and the Noether identities of fundamental matter as deduced in the framework of gauge theories of the Poincar\'e$\otimes U(1)$ group.
\end{abstract}

\maketitle

\section{Introduction}

The present work is based on our previous paper \cite{Tresguerres:2007ih}. There we studied jointly gravitation and electrodynamics in the form of a gauge theory of the Poincar\'e group times the internal group $U(1)$. Following the approach of Hehl et al. to gauge theories of gravity \cite{Hehl:1974cn}--\cite{Obukhov:2006ge}, we made use of a Lagrangian formalism to get the field equations and the Noether identities associated to the gauge symmetry, devoting special attention to energy conservation. This latter aspect of \cite{Tresguerres:2007ih}, where exchange between different forms of energy plays a central role, strongly suggests to look for a thermodynamic interpretation of the corresponding formulas, although this aim remains unattainable as only single matter particles are involved. For this reason, we are interested in extending similar energetic considerations to macroscopic matter in order to be able to construct an approach to thermodynamics compatible with gauge theories of gravity.

In this endeavor, our starting point is provided by the dynamical equations found for a particular form of fundamental matter, namely Dirac matter, with the help of the principle of invariance of the action under local Poincar\'e$\otimes U(1)$ transformations. Our main hypothesis is that the equations still hold for other forms of matter with the same $U(1)$, translational and Lorentz symmetry properties, and we assume that these are possessed by macroscopic matter. Accordingly, we consider that material media obey equations with a form which is known to us, also when we have to reinterpret several quantities involved in them --in particular the matter sources-- in order to give account of macroscopic features which are not present in the original formulation.

Moreover, a major alteration of the almost purely geometrical approach to physical reality characteristic for gauge theories occurs with the introduction of thermodynamic variables. Briefly exposed, regarding the latter ones we proceed as follows. From the original gauge theoretically defined matter energy current $\epsilon ^{\rm matt}$, we define a modified matter energy current $\epsilon ^{\rm u}$ with an energy flux component $q$ identified as heat flux, and a further component $\mathfrak{U}$ representing the internal energy content of a volume element. As a requirement of the transition to macroscopic matter \cite{Callen}, we postulate $\mathfrak{U}$ to depend, among others, on a new macroscopic variable $\mathfrak{s}$ with the meaning of the entropy content of an elementary volume. (Contrary to other authors \cite{Landau:1958}-\cite{Priou:1991}, we do not introduce an additional entropy flow variable.) The definition of temperature as the derivative of $\mathfrak{U}$ with respect to $\mathfrak{s}$ completes the set of fundamental thermal variables. We are going to prove that they satisfy the first and second laws of thermodynamics. In our approach, the energy and entropy forms, as much as the temperature function, are Lorentz invariants, as in Eckart's pioneering work \cite{Eckart:1940te}. There, as in our case, the first principle of thermodynamics is derived from the energy-momentum conservation law not as the zero component of this vector equation, but as a scalar equation.

The paper is organized as follows. In Sections II and III we present the gauge-theoretically derived field equations and Noether identities. After introducing in IV a necessary spacetime foliation, Section V is devoted to defining total energy and its various constitutive pieces, and to studying the corresponding conservation equations. In VI, explicit Lagrangians for electrodynamics and gravity are considered, while VII deals with some aspects of the energy-momentum of macroscopic matter. In Section VIII we argue on the most suitable way to include the features of material media in the dynamical equations. Lastly, the main results are presented in Section IX, where we deduce the laws of thermodynamics in two different scenarios. The paper ends with several final remarks and the conclusions.

\section{Field equations}

The results of \cite{Tresguerres:2007ih} relevant for the present paper are summarized in what follows with slight changes needed to replace the fundamental Dirac matter by macroscopic matter. Interested readers are referred to \cite{Tresguerres:2007ih} for technical details, in particular those concerning the handling of translations. A complementary study of the underlying geometry of dynamical spacetimes of Poincar\'e gauge theories can be found in Refs. \cite{Tresguerres:2002uh} and \cite{Tresguerres:2012nu}.

Our point of departure is a Lagrangian density 4-form
\begin{equation}
L=L(\,A\,,\vartheta ^\alpha\,,\Gamma ^{\alpha\beta}\,;F\,,T^\alpha\,,\,R^{\alpha\beta}\,;{\rm matter\hskip0.2cm variables}\,)\,,\label{totalLag}
\end{equation}
invariant under local Poincar\'e$\otimes U(1)$ symmetry. Its arguments, along with matter fields, are the following. On the one hand, we recognize the connection 1-forms of $U(1)$, of translations and of the Lorentz subgroup respectively: that is, the electromagnetic potential $A$, the (nonlinear) translational connections $\vartheta ^\alpha$ geometrically interpreted as tetrads, and the Lorentz connections $\Gamma ^{\alpha\beta}$ required to guarantee gauge covariance, being antisymmetric in their indices. On the other hand, further arguments are the covariantized derivatives of the preceding connections. The differential of the electromagnetic potential is the familiar electromagnetic field strength
\begin{equation}
F:= dA\,,\label{Fdef}
\end{equation}
and analogously, torsion \cite{Hehl:1995ue} defined as the covariant differential of tetrads
\begin{equation}
T^\alpha := D\,\vartheta ^\alpha = d\,\vartheta ^\alpha + \Gamma _\beta{}^\alpha\wedge\vartheta ^\beta\,,\label{torsiondef}
\end{equation}
together with the Lorentz curvature
\begin{equation}
R^{\alpha\beta} := d\,\Gamma ^{\alpha\beta} + \Gamma _\gamma{}^\beta\wedge \Gamma ^{\alpha\gamma}\,,\label{curvdef}
\end{equation}
play the role of the field strengths associated respectively to translations and to the Lorentz group. Lorentz indices are raised and lowered with the help of the constant Minkowski metric $o_{\alpha\beta}= diag(-+++)$.

The derivatives of (\ref{totalLag}) with respect to the connections $A$, $\vartheta ^\alpha $ and $\Gamma ^{\alpha\beta}$ are the electric four-current 3-form
\begin{equation}
J :={{\partial L}\over{\partial A}}\,,\label{definition03a}
\end{equation}
the total energy-momentum 3-form
\begin{equation}
\Pi _\alpha :={{\partial L}\over{\partial \vartheta ^\alpha}}\,,\label{definition03b}
\end{equation}
(including, as we will see, electrodynamic, gravitational and matter contributions), and the spin current\footnote{The definition of spin current given in Eq.(61) of Reference \cite{Tresguerres:2007ih} differs from the present one due to the fact that there we considered an internal structure for the tetrads, with a particular dependence on $\Gamma ^{\alpha\beta}$, giving rise to additional terms. The latter ones are not present when the internal structure of the tetrads is ignored, as is the case here.}
\begin{equation}
\tau _{\alpha\beta} :={{\partial L}\over{\partial \Gamma ^{\alpha\beta}}}\,.\label{definition03c}
\end{equation}
Finally, derivatives of (\ref{totalLag}) with respect to the field strengths (\ref{Fdef}), (\ref{torsiondef}) and (\ref{curvdef}) yield respectively the electromagnetic excitation 2-form
\begin{equation}
H:=-{{\partial L}\over{\partial F}}\,,\label{definition01}
\end{equation}
and its translative and Lorentzian analogs, defined as the excitation 2-forms
\begin{equation}
\quad H_\alpha :=-{{\partial L}\over{\partial T^\alpha}}\,,\quad H_{\alpha\beta}:=-\,{{\partial L}\over{\partial R^{\alpha\beta}}}\,.\label{definition02}
\end{equation}
With these definitions at hand, the principle of extremal action yields the field equations
\begin{eqnarray}
dH &=&J\,,\label{covfieldeq1} \\
DH_\alpha &=&\Pi _\alpha\,,\label{covfieldeq2}\\
DH_{\alpha\beta} +\vartheta _{[\alpha }\wedge H_{\beta ]}&=&\tau _{\alpha\beta}\,.\label{covfieldeq3}
\end{eqnarray}
As we will see below, suitable explicit Lagrangians uncover respectively (\ref{covfieldeq1}) as Maxwell's equations and (\ref{covfieldeq2}) as a generalized Einstein equation for gravity, whereas (\ref{covfieldeq3}) completes the scheme taking spin currents into account. Notice that Eqs. (\ref{covfieldeq1})--(\ref{covfieldeq3}) are explicitly Lorentz covariant\footnote{The covariant differentials in
(\ref{covfieldeq2}) and (\ref{covfieldeq3}) are defined as
$$DH_\alpha := dH_\alpha -\Gamma _\alpha{}^\beta\wedge H_\beta\,,$$
and
$$DH_{\alpha\beta} := dH_{\alpha\beta} -\Gamma _\alpha{}^\gamma\wedge H_{\gamma\beta}
-\Gamma _\beta{}^\gamma\wedge H_{\alpha\gamma}\,,$$ respectively.}. In addition, they are invariant with respect to translations as much as to $U(1)$ as a consequence of the (nonlinear) symmetry realization used in \cite{Tresguerres:2007ih}.

\section{Noether identities}

Following \cite{Hehl:1995ue}, we separate the total Lagrangian density 4-form (\ref{totalLag}) into three different pieces
\begin{equation}
L=L^{\rm matt}+L^{\rm em}+L^{\rm gr}\,,\label{Lagrangedecomp}
\end{equation}
consisting respectively in the matter contribution
\begin{equation}
L^{\rm matt} = L^{\rm matt}(\,\vartheta ^\alpha\,;{\rm matter\hskip0.2cm variables}\,)\,,\label{mattLagcontrib}
\end{equation}
(in the fundamental case, matter variables consisting of matter fields $\psi$ and of their covariant derivatives including connections $A$ and $\Gamma ^{\alpha\beta}$), together with the electromagnetic part $L^{\rm em}(\,\vartheta ^\alpha\,,\,F\,)\,$ and the gravitational Lagrangian $L^{\rm gr}(\,\vartheta ^\alpha\,,\,T^\alpha\,,\,R_\alpha{}^\beta\,)$. According to (\ref{Lagrangedecomp}), the energy-momentum 3-form (\ref{definition03b}) decomposes as
\begin{equation}
\Pi _\alpha =\Sigma ^{\rm matt}_\alpha +\Sigma ^{\rm em}_\alpha +E_\alpha\,,\label{momentdecomp}
\end{equation}
with the different terms in the right-hand side (rhs) defined respectively as
\begin{equation}
\Sigma ^{\rm matt}_\alpha :={{\partial L^{\rm matt}}\over{\partial \vartheta ^\alpha}}\,,\quad
\Sigma ^{\rm em}_\alpha :={{\partial L^{\rm em}}\over{\partial \vartheta ^\alpha}}\,,\quad
E_\alpha :={{\partial L^{\rm gr}}\over{\partial \vartheta ^\alpha}}\,.\label{momentdecompbis}
\end{equation}
Starting with the matter Lagrangian part $L^{\rm matt}\,$, let us derive the Noether type conservation equations for the matter currents associated to the different symmetries, that is
\begin{equation}
J={{\partial L^{\rm matt}}\over{\partial A}}\,,\quad
\Sigma ^{\rm matt}_\alpha = {{\partial L^{\rm matt}}\over{\partial \vartheta ^\alpha }}\,,\quad\tau _{\alpha\beta} = {{\partial L^{\rm matt}}\over{\partial \Gamma ^{\alpha\beta}}}\,.\label{mattcurrdefs}
\end{equation}
Provided the field equations (\ref{covfieldeq1})--(\ref{covfieldeq3}) are fulfilled, as much as the Euler-Lagrange equations for matter fields (non explicitly displayed here), from the invariance of $L^{\rm matt}$ under vertical (gauge) Poincar\'e $\otimes$ $U(1)$ transformations follow the conservation equations for both, the electric current
\begin{equation}
dJ =0\,,\label{elcurrcons}
\end{equation}
and the spin current
\begin{equation}
D\,\tau _{\alpha\beta} +\vartheta _{[\alpha}\wedge\Sigma ^{\rm matt}_{\beta ]}=0\,.\label{spincurrconserv}
\end{equation}
On the other hand, the Lie (lateral) displacement ${\it{l}}_{\bf x} L^{\rm matt}$ of the Lagrangian 4-form along an arbitrary vector field $X$ yields the identity
\begin{equation}
D\,\Sigma ^{\rm matt}_\alpha =(\,e_\alpha\rfloor T^\beta )\wedge\Sigma ^{\rm matt}_\beta +(\,e_\alpha\rfloor R^{\beta\gamma}\,)\wedge\tau _{\beta\gamma} +(\,e_\alpha\rfloor F\,)\wedge J\,,\label{sigmamattconserv}
\end{equation}
with the matter energy-momentum 3-form given by
\begin{equation}
\Sigma ^{\rm matt}_\alpha =-(\,e_\alpha\rfloor\overline{D\psi}\,)\,{{\partial L^{\rm matt}}\over{\partial d\overline{\psi}}} +{{\partial L^{\rm matt}}\over{\partial d\psi}}\,(\,e_\alpha\rfloor D\psi\,) + e_\alpha\rfloor L^{\rm matt}\label{sigmamatt}
\end{equation}
(for Dirac matter, and thus to be modified for the case of macroscopic matter). In the rhs of (\ref{sigmamattconserv}) we recognize, besides the proper Lorentz force 4-form in the extreme right, two additional terms with the same structure, built with the field strengths and the matter currents of translational and Lorentz symmetry respectively.

Next we apply the same treatment to the remaining constituents of (\ref{Lagrangedecomp}). The gauge invariance of the electromagnetic Lagrangian piece implies
\begin{equation}
\vartheta _{[\alpha}\wedge\Sigma ^{\rm em}_{\beta ]} =0\,,\label{Symem-emt}
\end{equation}
while in analogy to (\ref{sigmamattconserv}) we find
\begin{equation}
D\,\Sigma ^{\rm em}_\alpha =(\,e_\alpha\rfloor T^\beta )\wedge\Sigma ^{\rm em}_\beta -(\,e_\alpha\rfloor F\,)\wedge dH\,,\label{sigmaemconserv}
\end{equation}
being the electromagnetic energy-momentum
\begin{equation}
\Sigma ^{\rm em}_\alpha =(\,e_\alpha\rfloor F\,)\wedge H + e_\alpha\rfloor L^{\rm em}\,.\label{sigmaem}
\end{equation}
Finally, regarding the gravitational Lagrangian part, its gauge invariance yields
\begin{equation}
D\,\Bigl( DH_{\alpha\beta} +\vartheta _{[\alpha }\wedge H_{\beta ]}\,\Bigr) +\vartheta _{[\alpha}\wedge\Bigl( DH_{\beta ]} -E_{\beta ]}\,\Bigr)=0\,,\label{redund}
\end{equation}
(derivable alternatively from (\ref{spincurrconserv}) with (\ref{covfieldeq2}), (\ref{covfieldeq3}), (\ref{momentdecomp}) and (\ref{Symem-emt})), and the (\ref{sigmamattconserv}) and (\ref{sigmaemconserv})-- analogous equation reads
\begin{eqnarray}
&&D\,\Bigl( DH_\alpha -E_\alpha\,\Bigr) -(\,e_\alpha\rfloor T^\beta
)\wedge\Bigl( DH_\beta -E_\beta\,\Bigr)\nonumber\\
&&\hskip0.2cm -(\,e_\alpha\rfloor R^{\beta\gamma}\,)\wedge\Bigl( DH_{\beta\gamma}+\vartheta _{[\beta }\wedge H_{\gamma ]}\,\Bigr)=0\,,\label{ealphaconserv}
\end{eqnarray}
with the pure gravitational energy-momentum given by
\begin{eqnarray}
E_\alpha =(\,e_\alpha\rfloor T^\beta )\wedge H_\beta +(\,e_\alpha\rfloor R^{\beta\gamma}\,)\wedge H_{\beta\gamma} +e_\alpha\rfloor L^{\rm gr}\,.\label{ealpha}
\end{eqnarray}
Eq.(\ref{ealphaconserv}) is also redundant, being derivable from (\ref{sigmamattconserv}) and (\ref{sigmaemconserv}) together with the field equations (\ref{covfieldeq1})--(\ref{covfieldeq3}) and (\ref{momentdecomp}).

\section{Spacetime foliation}

\subsection{General formulas}

The definition of energy to be introduced in next section, as much as its subsequent thermodynamic treatment, rests on a foliation of spacetime involving a timelike vector field $u$ defined as follows. (For more details, see \cite{Tresguerres:2012nu}.) The foliation is induced by a 1-form $\omega = d\tau $ trivially satisfying the Frobenius' foliation condition $\omega\wedge d\omega =0$. The vector field $u$ relates to $d\tau$ through the condition $u\rfloor d\tau =1$ fixing its direction. This association of the vector $u$ with $\tau $, the latter being identified as {\it parametric time}, allows one to formalize time evolution of any physical quantity represented by a $p$-form $\alpha$ as its Lie derivative along $u$, that is
\begin{equation}
{\it{l}}_u\alpha :=\,d\,(u\rfloor\alpha\,) + u\rfloor d\alpha \,.\label{Liederdef}
\end{equation}
(Notice that the condition $u\rfloor d\tau =1$ itself defining $u$ in terms of $\tau$ means that ${\it l}_u\,\tau := u\rfloor d\tau =1$.) With respect to the direction of the time vector $u$, any $p$-form $\alpha$ decomposes into two constituents \cite{Hehl-and-Obukhov}, longitudinal and transversal to $u$ respectively, as
\begin{equation}
\alpha = d\tau\wedge\alpha _{\bot} +\underline{\alpha}\,,\label{foliat1}
\end{equation}
with the longitudinal piece
\begin{equation}
\alpha _{\bot} := u\rfloor\alpha\,,\label{long-part}
\end{equation}
consisting of the projection of $\alpha$ along $u$, and the transversal component
\begin{equation}
\underline{\alpha}:=
u\rfloor ( d\tau\wedge\alpha\,)\,,\label{trans-part}
\end{equation}
orthogonal to the former as a spatial projection.

The foliation of exterior derivatives of forms is performed in analogy to (\ref{foliat1}) as
\begin{equation}
d\,\alpha = d\tau\wedge\bigl(\,{\it{l}}_u\underline{\alpha} -\,\underline{d}\,\alpha _{\bot}\,\bigr) +\underline{d}\,\underline{\alpha }\,,\label{derivfoliat}
\end{equation}
with the longitudinal part expressed in terms of the Lie derivative (\ref{Liederdef}) and of the spatial differential $\underline{d}$. For its part, the Hodge dual (\ref{dualform}) of a $p$-form $\alpha$ decomposes as
\begin{equation}
{}^*\alpha =\,(-1)^p\, d\tau\wedge {}^{\#}\underline{\alpha} - {}^{\#}\alpha _{\bot}\,,\label{foliat2}
\end{equation}
being $^\#$ the Hodge dual operator in the three-dimensional spatial sheets.

\subsection{Foliation of tetrads}

Let us apply the general formulas (\ref{Liederdef})--(\ref{foliat2}) to the particular case of tetrads $\vartheta ^\alpha $, which, as universally coupling coframes \cite{Tresguerres:2007ih}, will play a significant role in what follows. Their dual vector basis $\{e_\alpha\}$ is defined by the condition
\begin{equation}
e_\alpha\rfloor \vartheta ^\beta = \delta _\alpha ^\beta\,.\label{dualitycond}
\end{equation}
When applied to tetrads, (\ref{foliat1}) reads
\begin{equation}
\vartheta ^\alpha = d\tau\,u^\alpha + \underline{\vartheta}^\alpha\,,\label{tetradfoliat}
\end{equation}
where the longitudinal piece
\begin{equation}
u^\alpha := u\rfloor\vartheta ^\alpha\label{fourvel}
\end{equation}
has the meaning of a four-velocity. In terms of it, the time vector $u$ can be expressed as $u =u^\alpha e_\alpha$, being the requirement for $u$ to be timelike fulfilled as
\begin{equation}
u_\alpha u^\alpha = -1\,.\label{form01}
\end{equation}
In terms of (\ref{fourvel}), let us define the projector
\begin{equation}
h_\alpha{}^\beta :=\delta _\alpha ^\beta + u_\alpha u^\beta\,.\label{form03}
\end{equation}
Replacing (\ref{tetradfoliat}) in (\ref{dualitycond}) and making use of (\ref{form03}) we find
\begin{equation}
e_\alpha\rfloor \Big(\,d\tau\,u^\beta + \underline{\vartheta}^\beta\,\Bigr) = \delta _\alpha ^\beta
=-u_\alpha u^\beta +h_\alpha{}^\beta \,.\label{dualitycondbis}
\end{equation}
implying
\begin{equation}
e_\alpha \rfloor d\tau = -\,u_\alpha\,,\label{form02}
\end{equation}
and
\begin{equation}
e_\alpha\rfloor \underline{\vartheta}^\beta = h_\alpha{}^\beta\,.\label{dualitycondbis}
\end{equation}
On the other hand, let us generalize the definition (\ref{Liederdef}) of Lie derivatives by considering covariant differentials instead of ordinary ones \cite{Hehl:1995ue}. In particular, we will make extensive use of the covariant Lie derivative of the tetrads, defined as
\begin{eqnarray}
{\cal \L\/}_u\vartheta ^\alpha &:=& D\left( u\rfloor\vartheta ^\alpha\right) + u\rfloor D\vartheta ^\alpha\nonumber\\
&=& D u^\alpha + T_{\bot}^\alpha
\,,\label{thetaLiederiv01}
\end{eqnarray}
where
\begin{equation}
{\cal \L\/}_u\vartheta ^\alpha = {\it{l}}_u\vartheta ^\alpha +{\Gamma _{\bot}}_\beta{}^\alpha\wedge\vartheta ^\beta\,,\label{thetaLiederiv02}
\end{equation}
with (\ref{thetaLiederiv01}) decomposing into the longitudinal and transversal pieces
\begin{eqnarray}
({\cal \L\/}_u\vartheta ^\alpha )_{\bot} &=& {\cal \L\/}_u u^\alpha\,,\label{thetaLiederiv03}\\
\underline{{\cal \L\/}_u\vartheta ^\alpha} &=& \underline{D} u^\alpha + T_{\bot}^\alpha\nonumber\\
&=& {\cal \L\/}_u\underline{\vartheta}^\alpha\,.\label{thetaLiederiv04}
\end{eqnarray}
For what follows, we also need complementary formulas concerning the foliation of the eta basis. Since they require more space, we introduce them in Appendix A.

\section{Definition and conservation of energy}

In Ref.\cite{Tresguerres:2007ih} we discussed the definition of the total energy current 3-form
\begin{equation}
\epsilon := -\left(\,u^\alpha\,\Pi _\alpha + Du^\alpha\wedge H_\alpha\,\right)\,.\label{energycurr}
\end{equation}
By rewriting it as
\begin{equation}
\epsilon =-d\left( u^\alpha H_\alpha\right) + u^\alpha \left( DH_\alpha -\Pi _\alpha \right)\,,\label{exactform01}
\end{equation}
and making use of (\ref{covfieldeq2}), we find that it reduces to an exact form
\begin{equation}
\epsilon =-d\left( u^\alpha H_\alpha\right)\,,\label{exactform02}
\end{equation}
automatically satisfying the continuity equation
\begin{equation}
d\,\epsilon =0\,.\label{energyconserv01}
\end{equation}
The interpretation of (\ref{energycurr}) as total energy, and thus of (\ref{energyconserv01}) as local conservation of total energy, becomes apparent with the help of (\ref{momentdecomp}). The energy (\ref{energycurr}) reveals to be the sum of three pieces
\begin{equation}
\epsilon =\epsilon ^{\rm matt}+\epsilon ^{\rm em}+\epsilon ^{\rm gr}\,,\label{energydec}
\end{equation}
defined respectively as
\begin{eqnarray}
\epsilon ^{\rm matt} &:=& -u^\alpha\,\Sigma ^{\rm matt}_\alpha\,,\label{mattenergy}\\
\epsilon ^{\rm em} &:=& -u^\alpha\,\Sigma ^{\rm em}_\alpha\,,\label{emenergy}\\
\epsilon ^{\rm gr} &:=& -\left(\,u^\alpha\,E_\alpha + D u^\alpha\wedge H_\alpha\,\right)\,.\label{grenergy}
\end{eqnarray}
On the other hand, decomposing (\ref{energycurr}) into its longitudinal and transversal components
\begin{equation}
\epsilon = d\tau\wedge\epsilon _{\bot} +\underline{\epsilon}\,,\label{energyfol01}
\end{equation}
the foliated form of the local energy conservation equation (\ref{energyconserv01}) reads
\begin{equation}
{\it l}_u\,\underline{\epsilon}-\underline{d}\,\epsilon _{\bot}=0\,,\label{conteq}
\end{equation}
showing (when integrated) that the rate of increase of the energy $\underline{\epsilon}$ contained in a small volume equals the amount of energy flowing into the volume over its boundary surface as the result of the balance of inflow and outflow of the energy flux $\epsilon _{\bot}$ crossing through the closed surface.

Conservation of total energy is the result of exchanges between the different forms of energy. Let us write the continuity equations of the different pieces (\ref{mattenergy})--(\ref{grenergy}). As we will see immediately, in all these equations, when considered separately, sources and sinks of energy are involved, reflecting the fact that, inside the small volume considered, energy is produced or consumed, wether on account of work or of any other manifestation of energy. These terms only cancel out when all forms of energy are considered together, that is, in (\ref{energyconserv01}) with (\ref{energydec}).

Regarding the matter contribution to energy (\ref{mattenergy}), using (\ref{sigmamattconserv}) we find
\begin{equation}
d\,\epsilon ^{\rm matt} = -{\cal \L\/}_u\,\vartheta ^\alpha\wedge\Sigma ^{\rm matt}_\alpha -R_{\bot}^{\alpha\beta}\wedge\tau _{\alpha\beta} -F_{\bot}\wedge J\,.\label{mattender}
\end{equation}
The interpretation of this conservation equation when its validity is extended to macroscopic matter constitutes the main task of the present work. Actually, Eq. (\ref{mattender}) provides the basis for our approach to thermodynamics.

In analogy to (\ref{mattender}), definition (\ref{emenergy}) of electromagnetic energy  with (\ref{sigmaemconserv}) yields the Poynting equation
\begin{equation}
d\,\epsilon ^{\rm em} = -{\cal \L\/}_u\,\vartheta ^\alpha\wedge\Sigma ^{\rm em}_\alpha + F_{\bot}\wedge dH\,,\label{emender}
\end{equation}
generalized to take into account spacetime as defined in Poincar\'e gauge theories. In (\ref{emender}), the energy flux (or intensity of flowing energy) is represented by the Poynting 2-form $\epsilon ^{\rm em}_{\bot}$, and the last term in the rhs is related to Joule's heat. Finally, from the gravitational energy definition (\ref{grenergy}) with (\ref{ealphaconserv}) we get
\begin{eqnarray}
d\,\epsilon ^{\rm gr} &:=& -{\cal \L\/}_u\,\vartheta ^\alpha\wedge\left(\,E_\alpha -DH_\alpha\right)\nonumber\\
&&+R_{\bot}^{\alpha\beta}\wedge \left(\,DH_{\alpha\beta} +\vartheta _{[\alpha }\wedge H_{\beta ]}\right)\,.\label{grender}
\end{eqnarray}
The field equations (\ref{covfieldeq1})--(\ref{covfieldeq3}) guarantee that the sum of (\ref{mattender}), (\ref{emender}) and (\ref{grender}) is conserved, in agreement with (\ref{energyconserv01}).

\section{Electrodynamical and gravitational Lagrangians}

In the present Section we introduce explicit Lagrangian pieces (\ref{Lagrangedecomp}) describing electrodynamics and gravity. We do so in order to calculate in particular the excitations defined in (\ref{definition01}) and (\ref{definition02}), which extend to the macroscopic arena without alterations, as will be discussed in Section VIII. We also derive the electromagnetic and gravitational energy-momentum  contributions to (\ref{momentdecomp}) as defined in (\ref{momentdecompbis}), and the corresponding energies (\ref{emenergy}) and (\ref{grenergy}). The form found for (\ref{emenergy}), namely (\ref{explemen1}), and in particular that of its transversal part (\ref{emendh}), provides us with a criterion to choose the way to extend the {\it microscopic} fundamental equations to macroscopic material media. (See Section VIII.)

\subsection{Electrodynamics}

In the context of fundamental matter in vacuum, we consider the Maxwell Lagrangian
\begin{equation}
L^{\rm em}=-{1\over 2}\,F\wedge\,^*F\,.\label{emlagrang1}
\end{equation}
From it follows a field equation of the form (\ref{covfieldeq1}) where the excitation (\ref{definition01}) is given by the Maxwell-Lorentz electromagnetic spacetime relation
\begin{equation}
H={}^*F\,,\label{emmom}
\end{equation}
involving (\ref{Fdef}), which identically satisfies
\begin{equation}
dF =0\,.\label{vanfder}
\end{equation}
Eqs. (\ref{covfieldeq1}) and (\ref{vanfder}) complete the set of Maxwell's equations for fundamental matter in vacuum.

On the other hand, the electromagnetic part (\ref{sigmaem}) of energy-momentum derived from the explicit Lagrangian (\ref{emlagrang1}) reads
\begin{equation}
\Sigma ^{\rm em}_\alpha = {1\over 2}\,\left[\,\left( e_\alpha\rfloor F\right)\wedge H -F\wedge\left( e_\alpha\rfloor H\right)\,\right]\,,\label{emenergymom}
\end{equation}
so that (\ref{emenergy}) becomes
\begin{equation}
\epsilon ^{\rm em} = -{1\over 2}\,\bigl(\,F_{\bot}\wedge H -F\wedge H_{\bot}\,\bigr)\,,\label{explemen1}
\end{equation}
obeying Eq.(\ref{emender}). The transversal component $\underline{\epsilon}^{\rm em}$ of the electromagnetic energy current 3-form (\ref{explemen1}) is the energy 3-form representing the amount of electric and magnetic energy contained in a small volume, and the longitudinal part $\epsilon ^{\rm em}_{\bot}$  is the energy flux or Poynting 2-form.

\subsection{Gravity}

For the gravitational action, we consider a quite general Lagrangian density taken from Ref. \cite{Obukhov:2006ge}, including a Hilbert-Einstein term with cosmological constant, plus additional contributions quadratic in the Lorentz-irreducible pieces of torsion and curvature as established by McCrea \cite{Hehl:1995ue} \cite{McCrea:1992wa}. The gravitational Lagrangian reads
\begin{eqnarray}
L^{\rm gr}&=&{1\over{\kappa}}\,\left(\,\,{a_0\over
2}\,\,R^{\alpha\beta}\wedge\eta_{\alpha\beta}
-\Lambda\,\eta\,\right)\nonumber\\
&&-{1\over 2}\,\,T^\alpha\wedge
\left(\sum_{I=1}^{3}{{a_{I}}\over{\kappa}}\,\,{}^{*(I)}
T_\alpha\right)\nonumber\\
&&-{1\over 2}\,\,R^{\alpha\beta}\wedge\left(\sum_{I=1}^{6}b_{I}\,\,
{}^{*(I)}R_{\alpha\beta}\right)\,,\label{gravlagr}
\end{eqnarray}
with $\kappa$ as the gravitational constant, and $a_0$, $a_{I}$, $b_{I}$ as dimensionless constants. From (\ref{gravlagr}) we calculate the translational and Lorentz excitations (\ref{definition02}) to be respectively
\begin{eqnarray}
H_\alpha &=& \sum_{I=1}^{3}{{a_{I}}\over{\kappa}}\,\,{}^{*(I)}
T_\alpha\,,\label{torsmom}\\
H_{\alpha\beta}&=&-{a_0\over{2\kappa}}\,\eta_{\alpha\beta} +\sum_{I=1}^{6}b_{I}\,\,
{}^{*(I)}R_{\alpha\beta}\,,\label{curvmom}
\end{eqnarray}
and we find the pure gravitational contribution (\ref{ealpha}) to the energy-momentum
\begin{eqnarray}
E_\alpha &=& {a_0\over {4\kappa}}\,e_\alpha\rfloor \left(\,R^{\beta\gamma}\wedge\eta_{\beta\gamma}\,\right)-{\Lambda\over{\kappa}}\,\eta _\alpha\nonumber\\
&&+{1\over 2}\,\left[\,\left( e_\alpha\rfloor T^\beta\right)\wedge H_\beta  -T^\beta\wedge\left( e_\alpha\rfloor H_\beta \right)\,\right]\nonumber\\
&&+{1\over 2}\,\left[\,\left( e_\alpha\rfloor R^{\beta\gamma}\right)\wedge H_{\beta\gamma}  -R^{\beta\gamma}\wedge\left( e_\alpha\rfloor H_{\beta\gamma}\right)\,\right]\,.\nonumber\\
\label{gravenergymom}
\end{eqnarray}
(Notice the resemblance between (\ref{gravenergymom}) and (\ref{emenergymom}).) The gauge-theoretical equations (\ref{covfieldeq2}) with (\ref{gravenergymom}) and (\ref{momentdecomp}) constitute a generalization of Einstein's equations. Actually, for $a_0=1\,$, $a_{I}=0\,$, $b_{I}=0\,$ and vanishing torsion, (\ref{gravenergymom}) reduces to
\begin{equation}
E_\alpha = {1\over{\kappa}}\,\left(\,\,{1\over 2}\,\,R^{\beta\gamma}\wedge\eta_{\beta\gamma\alpha}
-\Lambda\,\eta _\alpha\,\right)\,,\label{H-Egravenergymom}
\end{equation}
which is simply an exterior calculus reformulation of Einstein's tensor plus a cosmological constant term. Using the general expression (\ref{gravenergymom}), we calculate the gravitational energy (\ref{grenergy}) to be
\begin{eqnarray}
\epsilon ^{\rm gr} &=& -{a_0\over {4\kappa}}\,\bigl(\,R^{\alpha\beta}\wedge\eta_{\alpha\beta}\,\bigr)_{\bot}+{\Lambda\over{\kappa}}\,u^\alpha\eta _\alpha\nonumber\\
&&-{1\over 2}\,\bigl(\,T_{\bot}^\alpha\wedge H_\alpha  -T^\alpha\wedge H_{{\bot}\alpha}\,\bigr)\nonumber\\
&&-{1\over 2}\,\bigl(\,R_{\bot}^{\alpha\beta}\wedge H_{\alpha\beta}  -R^{\alpha\beta}\wedge H_{{\bot}\alpha\beta}\,\bigr)\nonumber\\
&&-D u^\alpha\wedge H_\alpha\,,\label{explgren}
\end{eqnarray}
(compare with (\ref{explemen1})), obeying Eq.(\ref{grender}).

\section{Energy-momentum 3-form of macroscopic matter}

Contrarily to the former cases of electromagnetism and gravity, we do not propose a Lagrangian for macroscopic matter. Instead, we focus our attention on the matter energy-momentum 3-form $\Sigma ^{\rm matt}_\alpha $, for which we postulate the dynamical equation (\ref{sigmamattconserv}), and any other in which it appears, to hold macroscopically. The energy-momentum  (\ref{sigmamatt}) found for Dirac matter does not play any role when considering macroscopic systems. The description of each kind of material medium requires the construction of a suitably chosen energy-momentum  3-form adapted to it. In the present Section we merely present a useful decomposition applicable to any $\Sigma ^{\rm matt}_\alpha$, and we consider the form of the simplest of all mechanic energy-momentum  contributions, namely that due to pressure, which we explicitly separate from the whole macroscopic matter energy-momentum. By using projectors (\ref{form03}) and definition (\ref{mattenergy}), we find
\begin{eqnarray}
\Sigma ^{\rm matt}_\alpha &&\equiv ( -u_\alpha u^\beta + h_\alpha{}^\beta ) \Sigma ^{\rm matt}_\beta\nonumber\\
&&=: u_\alpha\,\epsilon ^{\rm matt} +\widetilde{\Sigma}^{\rm matt}_\alpha\,,\label{enmom02}
\end{eqnarray}
making apparent the pure energy content of energy-momentum . On the other hand, to give account of pressure, we separate the pressure term from an energy-momentum  3-form as
\begin{eqnarray}
\Sigma ^{\rm matt}_\alpha &=& p\,h_\alpha{}^\beta\,\eta _\beta +\Sigma ^{\rm undef}_\alpha\nonumber\\
&=&-d\tau\wedge p\,\overline{\eta}_\alpha +\Sigma ^{\rm undef}_\alpha\,,\label{enmom01}
\end{eqnarray}
with $\overline{\eta}_\alpha$ as defined in (\ref{3deta07}), while $\Sigma ^{\rm undef}_\alpha $ is left undefined. By decomposing (\ref{enmom01}) according to (\ref{enmom02}), we get
\begin{equation}
\Sigma ^{\rm matt}_\alpha = u_\alpha\,\epsilon ^{\rm matt} -d\tau\wedge p\,\overline{\eta}_\alpha +\widetilde{\Sigma}^{\rm undef}_\alpha\,.\label{enmom03}
\end{equation}
The piece $\widetilde{\Sigma}^{\rm undef}_\alpha $ present in (\ref{enmom03}) after the separation of the energy term can be chosen in different manners to describe, as the case may be, viscosity, elasticity, plasticity, etc. Actually, (\ref{enmom03}) resembles the energy-momentum  3-form of a fluid plus additional contributions responsible for different mechanic features.

Notice that, being (\ref{sigmamattconserv}) a dynamical equation of the form
\begin{equation}
D\,\Sigma ^{\rm matt}_\alpha = f_\alpha\,,\label{force01}
\end{equation}
where the 4-form $f_\alpha$ is a generalized Lorentz force, by replacing (\ref{enmom03}) in it, we get (at least formally) an extended Navier-Stokes equation.

\section{Electrodynamic equations in material media}

Looking for a general criterion about the most suitable procedure to include phenomenological matter in the fundamental equations, let us examine in particular electromagnetism in order to find out how to generalize (\ref{covfieldeq1}) as much as (\ref{emender}) in such a manner that they become applicable macroscopically while preserving their form. As a matter of fact, Maxwell's equations in matter admit two alternative formulations, depending on how the electric and magnetic properties of material media are taken into account \cite{Hehl-and-Obukhov} \cite{Obukhov:2003cc}. Actually, polarization and magnetization can be described, in seemingly equivalent ways, either as due to modifications of the electromagnetic excitations $H$ or as the result of the existence inside such materials of generalized currents $J$ including both, free and bound contributions. With the latter approach in mind, we define the total current density $J^{\rm tot}$ as the sum of a current $J^{\rm free}$ of free charge and a matter-bounded contribution $J^{\rm matt}$ characteristic for the medium, that is
\begin{equation}
J^{\rm tot} = J^{\rm free} + J^{\rm matt}\,,\label{totcurr01}
\end{equation}
with the assumption that they are conserved separately as
\begin{equation}
dJ^{\rm free}=0\,,\qquad dJ^{\rm matt}=0\,,\label{totcurrconserv}
\end{equation}
so that, although both types of charge can coexist, no exchange occurs between them. From the second conservation condition in (\ref{totcurrconserv}), we infer the existence of an independent excitation 2-form, which we denote as $H^{\rm matt}$, such that
\begin{equation}
J^{\rm matt}= -dH^{\rm matt}\,.\label{indepexcits}
\end{equation}
For the longitudinal and transversal pieces of $H^{\rm matt}$ we introduce the notation
\begin{equation}
H^{\rm matt}= -d\tau\wedge M + P\,,\label{matexcit01}
\end{equation}
where $M$ is the magnetization 1-form and $P$ the polarization 2-form.

The extension of Maxwell's equations (\ref{covfieldeq1}) to include the contribution (\ref{indepexcits}) of the material medium without altering their form can then be performed in any of the alternative ways mentioned above. Let us define
\begin{equation}
H^{\rm bare} :={}^*F\,,\label{macMax05}
\end{equation}
(where we call {\it bare fields} the fields in vacuum) in analogy to the Maxwell-Lorentz spacetime relation (\ref{emmom}). Then, according to the first procedure, consisting in considering the electromagnetic effects of the medium as due to a modification of the electromagnetic excitations, the latter ones $H$ as much as $J$ in (\ref{covfieldeq1}) are to be understood respectively as
\begin{equation}
H = H^{\rm tot} := H^{\rm bare} +H^{\rm matt}\quad{\rm and}\quad J= J^{\rm free}\,,\label{secondcase}
\end{equation}
while in the second case such effects are characterized in terms of bounded currents, so that the same equation (\ref{covfieldeq1}) is to be read taking in it now
\begin{equation}
H = H^{\rm bare}\quad{\rm and}\quad J = J^{\rm tot} := J^{\rm free} - dH^{\rm matt}\,.\label{firstcase}
\end{equation}
Let us show that, despite appearances, both formulations are not trivially interchangeable. Actually, only one of them can be easily adjusted to our program of generalizing the {\it microscopic} formulas (\ref{mattender}) and (\ref{emender}) to include the contributions of the medium. Our main argument to decide in favor of one of both alternatives (in the present context) is that the electromagnetic energy (\ref{explemen1}) is different in each case, in such a way that, for arbitrary $P$ and $M$, Eq.(\ref{emender}) is compatible with only one of the possible choices.

Making use of (\ref{foliat1}), we decompose the electromagnetic excitation 2-form $H$, the electromagnetic field strength 2-form $F$ and the current $J$ of Maxwell's equations (\ref{covfieldeq1}) and (\ref{vanfder}) as
\begin{eqnarray}
H &=& d\tau\wedge {\cal H} + {\cal D}\,,\label{Max01}\\
F &=& -d\tau\wedge E + B\,,\label{Max02}\\
J &=& -d\tau\wedge j + \rho\,.\label{Max03}
\end{eqnarray}
Accordingly, the foliation of (\ref{covfieldeq1}) yields
\begin{eqnarray}
{\it{l}}_u {\cal D} -\underline{d}\,{\cal H} &=& -j\,,\label{Max07}\\
\underline{d}\,{\cal D}&=& \rho\,.\label{Max08}
\end{eqnarray}
and that of (\ref{vanfder}) gives rise to
\begin{eqnarray}
{\it{l}}_u B +\underline{d}\,E &=& 0\,,\label{Max09}\\
\underline{d}\,B &=& 0\,.\label{Max10}
\end{eqnarray}
In Eqs. (\ref{Max07})--(\ref{Max10}) we do not prejudge which of both interpretations is to be given to the different fields. In order to decide, we express (\ref{macMax05}) in terms of the Hodge dual (\ref{foliat2}) of (\ref{Max02})
\begin{equation}
^*F = d\tau\wedge{}^\#B + {}^\#E\,.\label{Max04}
\end{equation}
So we see that (\ref{secondcase}) corresponds to the choice
\begin{equation}
{\cal D} ={}^\#E +P\,,\quad {\cal H}={}^\#B -M\,,\quad J=J^{\rm free}\,,\label{elmagexcits02}
\end{equation}
in the Maxwell equations (\ref{Max07})--(\ref{Max10}), with
\begin{equation}
J^{\rm free}=  -d\tau\wedge j^{\rm free} + \rho ^{\rm free}\,,\label{freecurr}
\end{equation}
while (\ref{firstcase}) gives rise to
\begin{equation}
{\cal D} ={}^\#E\,,\quad {\cal H}={}^\#B\,,\quad J=J^{\rm tot}\,,\label{elmagexcits01}
\end{equation}
being
\begin{equation}
J^{\rm tot}=  -d\tau\wedge ( j^{\rm free} +{\it{l}}_u P +\underline{d}\,M\,) + (\rho ^{\rm free}-\underline{d}\,P\,)\,,\label{totcurr}
\end{equation}
as calculated from (\ref{totcurr01}) with (\ref{indepexcits}) and (\ref{matexcit01}). Now, in order to check the compatibility either of (\ref{elmagexcits02}) or (\ref{elmagexcits01}) with (\ref{emender}), we add (\ref{Max07}) and (\ref{Max09}) to each other, respectively multiplied by $E$ and ${\cal H}$, to get
\begin{equation}
E\wedge{\it{l}}_u {\cal D} + {\it{l}}_u B\wedge{\cal H} +\underline{d}\,(E\wedge {\cal H}) = -E\wedge j\,,\label{Poynting01}
\end{equation}
and on the other hand, we rewrite the transversal part of (\ref{explemen1}) as
\begin{equation}
\underline{\epsilon}^{\rm em} ={1\over 2}\,( E\wedge{\cal D} + B\wedge {\cal H}\,)\,.\label{emendh}
\end{equation}
We can see that, in general, for nonspecified $P$ and $M$, the step from (\ref{Poynting01}) to (\ref{emender}) with $\epsilon ^{\rm em}$ given by (\ref{emendh}) is only possible with the choice (\ref{elmagexcits01}) for the excitations. Indeed, notice that the first term in the rhs of (\ref{emender}) has its origin in the relation
\begin{eqnarray}
&&{\it{l}}_u \underline{\epsilon}^{\rm em} := {\it{l}}_u\,{1\over 2}\left( E\wedge{}^\#E + B\wedge {}^\#B\,\right)\nonumber\\
&&\hskip1.0cm \equiv E\wedge {\it{l}}_u {}^{\#}E + {\it{l}}_u B\wedge {}^{\#}B -({\cal \L\/}_u\,\vartheta ^\alpha\wedge\Sigma ^{\rm em}_\alpha )_{\bot}\,,\nonumber\\
\label{ident02}
\end{eqnarray}
derived with the help of the identities
\begin{eqnarray}
{\it{l}}_u {}^{\#}E &\equiv &\,{}^{\#}\Bigl(\,{\it{l}}_u E -{\cal \L\/}_u\underline{\vartheta}^\alpha\, e_\alpha\rfloor E\,\Bigr) +{\cal \L\/}_u\underline{\vartheta}^\alpha\wedge\left( e_\alpha\rfloor {}^{\#}E\,\right)\,,\nonumber\\
\label{formula01}\\
{\it{l}}_u {}^{\#}B &\equiv &\,{}^{\#}\Bigl(\,{\it{l}}_u B -{\cal \L\/}_u\underline{\vartheta}^\alpha\wedge e_\alpha\rfloor B\,\Bigr) +{\cal \L\/}_u\underline{\vartheta}^\alpha\wedge\left( e_\alpha\rfloor {}^{\#}B\,\right)\,.\nonumber\\
\label{formula02}
\end{eqnarray}
(Compare with (\ref{dualvar}).) Thus, although (\ref{Poynting01}) holds in both approaches, it only can be brought to the form (\ref{emender}) within the scope of choice (\ref{elmagexcits01}), or equivalently of (\ref{firstcase}), the latter thus revealing to be necessary in order to guarantee the general applicability of the fundamental formulas found for microscopic matter. Accordingly, we choose option (\ref{firstcase}), which in practice means that, in order to apply the original formula (\ref{covfieldeq1}) of the fundamental approach, we have to keep in it the excitation $H =H^{\rm bare}={}^*F$ built from bare fields, and to include all contributions of the medium in the matter current by replacing $J$ by $J^{\rm tot}=J -dH^{\rm matt}$, where the new $J$ in $J^{\rm tot}$ is understood to be $J^{\rm free}$.

In the following, we generalize this criterion of strict separation between bare electromagnetic fields (say radiation in vacuum) and matter, in such a way that it also applies to the gravitational case. So, in all field equations and Noether identities established in Sections II and III, we have to leave untouched the excitations $H$, $H_\alpha$, $H_{\alpha\beta}$ built from bare fields as in Section VI, while modifying the matter currents $J$, $\Sigma ^{\rm matt}_\alpha $, $\tau _{\alpha\beta}$. The matter contributions separated from bare fields will enter $\epsilon ^{\rm matt}$ and thus $\epsilon ^{\rm u}$ as defined in Section IX, so that they will play a role in the thermodynamic relations to be established there.

\section{Deduction of the laws of thermodynamics}

\subsection{First approach, in an electromagnetic medium}

In view of the discussion of previous section, we identify $H$ with $H^{\rm bare}$ and, in order to adapt Eq.(\ref{mattender}) to a macroscopic medium with electromagnetic properties, we replace in it (as everywhere) $J$ by $J^{\rm tot}$, that is
\begin{equation}
d\,\epsilon ^{\rm matt} = -{\cal \L\/}_u\,\vartheta ^\alpha\wedge\Sigma ^{\rm matt}_\alpha -R_{\bot}^{\alpha\beta}\wedge\tau _{\alpha\beta} -F_{\bot}\wedge J^{\rm tot}\,.\label{emmattender}
\end{equation}
Taking into account the explicit form (\ref{totcurr}) of $J^{\rm tot}$, we find that (\ref{emmattender}) can be rewritten as
\begin{eqnarray}
&&\mkern-60mu d\,\bigl(\,\epsilon ^{\rm matt} +F\wedge M\,\bigr)\nonumber\\
&&= -{\cal \L\/}_u\,\vartheta ^\alpha\wedge\Sigma ^{\rm matt}_\alpha -R_{\bot}^{\alpha\beta}\wedge\tau _{\alpha\beta}-F_{\bot}\wedge J\nonumber\\
&&\quad + d\tau\wedge\Bigl\{ -F_{\bot}\wedge{\it l}_u P +\underline{F}\wedge{\it l}_u M\Bigr\}\,,\label{diff02}
\end{eqnarray}
where we use simply $J$ instead of $J^{\rm free}$. Let us define the modified matter energy current in the left-hand side (lhs) of (\ref{diff02}) as
\begin{equation}
\epsilon ^{\rm u} := \epsilon ^{\rm matt} + F\wedge M\,.\label{intenergycurr01}
\end{equation}
Then, from (\ref{diff02}) and (\ref{intenergycurr01}) and using the notation (\ref{Max02}) for $F$, we find the more explicit version of (\ref{diff02})
\begin{eqnarray}
d\,\epsilon ^{\rm u} &=&d\tau\wedge\Bigl\{{\Sigma _{\alpha}}_{\bot}^{\rm matt}\wedge{\cal \L\/}_u\underline{\vartheta}^\alpha -\underline{\Sigma}^{\rm matt}_\alpha\,{\cal \L\/}_u u^\alpha +R_{\bot}^{\alpha\beta}\wedge{\tau _{\alpha\beta}}_{\bot}\nonumber\\
&&\hskip1.5cm +E\wedge j +E\wedge{\it l}_u P +B\wedge{\it l}_u M \Bigr\}\,,\label{diff01}
\end{eqnarray}
where we recognize in the rhs, among other forms of energy, the electric and magnetic work contributions $E\wedge{\it l}_u P$ and $B\wedge{\it l}_u M$ respectively. Let us now decompose (\ref{intenergycurr01}) foliating it according to (\ref{foliat1}) and introducing a suitable notation for the longitudinal and transversal pieces, namely
\begin{eqnarray}
\epsilon ^{\rm u} &=& d\tau\wedge\epsilon ^{\rm u}_{\bot} + \underline{\epsilon}^{\rm u}\nonumber\\
&=:& d\tau\wedge q + \mathfrak{U}\,.\label{intenergycurr02}
\end{eqnarray}
As we are going to justify in the following (in view of the equations satisfied by these quantities), $q$ will play the role of the heat flux 2-form and $\mathfrak{U}$ that of the internal energy 3-form. From (\ref{intenergycurr02}) with (\ref{derivfoliat}) we get
\begin{equation}
d\,\epsilon ^{\rm u}= d\tau\wedge\left(\,{\it l}_u\,\mathfrak{U} - \underline{d}\,q\,\right)\,.\label{energycurrder01}
\end{equation}
At this point, we claim as a characteristic of macroscopic matter systems \cite{Callen} the dependence of the internal energy 3-form $\mathfrak{U}$ on a certain new quantity $\mathfrak{s}$ --the entropy-- which we take to be a spatial 3-form (representing the amount of entropy contained in an elementary volume). Eq.(\ref{secondlaw}) to be found below confirms {\it a posteriori} that $\mathfrak{s}$ actually behaves as expected for entropy. Moreover, the structure of (\ref{diff01}) suggests to promote a shift towards a fully phenomenological approach by considering $\mathfrak{U}$ to possess \cite{Callen} the following general functional dependence
\begin{equation}
\mathfrak{U} = \mathfrak{U}\,(\mathfrak{s}\,,P\,,M\,,\underline{\vartheta}^\alpha \,, u^\alpha\,)\,.\label{uargs}
\end{equation}
In (\ref{uargs}), as in the matter Lagrangian piece (\ref{mattLagcontrib}), tetrads are still taken as arguments of $\mathfrak{U}$ while new variables replace the fundamental matter fields $\psi$ and their covariant derivatives $D\psi$. Connections involved in the derivatives $D\psi$ are thus excluded together with the fields. Besides the new entropy variable and the polarization and magnetization of the medium (induced by external fields), we find the components
(\ref{tetradfoliat}) of the tetrads in terms of which the volume 3-form (\ref{3deta06}) with (\ref{3deta09}) is defined. Accordingly, the Lie derivative of (\ref{uargs}) present in (\ref{energycurrder01}) takes the form
\begin{eqnarray}
{\it l}_u\,\mathfrak{U} &=& {{\partial\mathfrak{U}}\over{\partial\mathfrak{s}}}\,{\it l}_u\mathfrak{s}
+{{\partial\mathfrak{U}}\over{\partial P}}\wedge{\it l}_u P
+{{\partial\mathfrak{U}}\over{\partial M}}\wedge{\it l}_u M\nonumber\\
&&+{{\partial\mathfrak{U}}\over{\partial\underline{\vartheta}^\alpha}}\wedge{\it l}_u\underline{\vartheta}^\alpha
+{{\partial\mathfrak{U}}\over{\partial u^\alpha}}\,{\it l}_u u^\alpha\,,\label{uLiederiv01}
\end{eqnarray}
where we identify the derivatives \cite{Callen} as
\begin{eqnarray}
&& {{\partial\mathfrak{U}}\over{\partial\mathfrak{s}}} =T\,,\quad
{{\partial\mathfrak{U}}\over{\partial P}} =E\,,\quad
{{\partial\mathfrak{U}}\over{\partial M}} =B\,,\label{Uder01}\\
&&{{\partial\mathfrak{U}}\over{\partial\underline{\vartheta}^\alpha}}={\Sigma _{\alpha}}_{\bot}^{\rm matt}\,,\quad
{{\partial\mathfrak{U}}\over{\partial u^\alpha}}=-\underline{\Sigma}^{\rm matt}_\alpha\,.\label{Uder02}
\end{eqnarray}
Let us call attention to the temperature defined in (\ref{Uder01}) as the derivative of the internal energy with respect to the entropy. On the other hand, a plausibility argument to justify the identifications we make in (\ref{Uder02}) can be found in Appendix B. Replacing (\ref{Uder01})--(\ref{Uder02}) in (\ref{uLiederiv01}) we get
\begin{eqnarray}
{\it l}_u\,\mathfrak{U} &=& T\,{\it l}_u\mathfrak{s}
+E\wedge{\it l}_u P
+B\wedge{\it l}_u M\nonumber\\
&&+{\Sigma _{\alpha}}_{\bot}^{\rm matt}\wedge{\it l}_u\underline{\vartheta}^\alpha
-\underline{\Sigma}^{\rm matt}_\alpha\,{\it l}_u u^\alpha\,.\label{uLiederiv02}
\end{eqnarray}
In order to rearrange the non explicitly invariant terms in (\ref{uLiederiv02}) to get invariant expressions, we replace the ordinary Lie derivatives by covariant Lie derivatives of the form (\ref{thetaLiederiv02}), so that the last terms in (\ref{uLiederiv02}) become
\begin{eqnarray}
{\Sigma _{\alpha}}_{\bot}^{\rm matt}\wedge{\it l}_u\underline{\vartheta}^\alpha -\underline{\Sigma}^{\rm matt}_\alpha\,{\it l}_u u^\alpha
&\equiv& {\Sigma _{\alpha}}_{\bot}^{\rm matt}\wedge{\cal \L\/}_u\underline{\vartheta}^\alpha
-\underline{\Sigma}^{\rm matt}_\alpha\,{\cal \L\/}_u u^\alpha\nonumber\\
&&+\Gamma _{\bot}^{\alpha\beta}\bigl(\,\vartheta _{[\alpha}\wedge\Sigma ^{\rm matt}_{\beta ]}\bigr)_{\bot}\,.\label{identity01}
\end{eqnarray}
Replacing (\ref{identity01}) in (\ref{uLiederiv02}) we finally arrive at
\begin{eqnarray}
{\it l}_u\,\mathfrak{U} &=& T\,{\it l}_u\mathfrak{s} +E\wedge{\it l}_u P +B\wedge{\it l}_u M\nonumber\\
&&+{\Sigma _{\alpha}}_{\bot}^{\rm matt}\wedge{\cal \L\/}_u\underline{\vartheta}^\alpha -\underline{\Sigma}^{\rm matt}_\alpha\,{\cal \L\/}_u u^\alpha\nonumber\\
&&+\Gamma _{\bot}^{\alpha\beta}\bigl(\,\vartheta _{[\alpha}\wedge\Sigma ^{\rm matt}_{\beta ]}\bigr)_{\bot}\,.\label{uLiederiv03}
\end{eqnarray}
In the rhs of (\ref{uLiederiv03}), the term containing explicitly the Lorentz connection is obviously noninvariant. Its emergence is due to an inherent limitation of the phenomenological approach, namely the absence of explicit dependence of $\mathfrak{U}$ on fundamental matter fields and their derivatives, together wit connections. Indeed, provided matter fields $\psi$ with derivatives $d\psi$ were present, connections were required to define covariant derivatives preserving local symmetry. However, in the phenomenological case, $\mathfrak{U}$ depends neither on $\psi$ nor on $d\psi$, so that (since $d\psi$ and connections need each other) it cannot give rise to invariant expressions, either one takes it or not to depend on the connections. The noninvariant term in (\ref{uLiederiv03}), reflecting the lack of invariance of the terms in the lhs of (\ref{identity01}), will be dragged to equations (\ref{energycurrder02}) and (\ref{secondlaw}) below. (We will find a similar situation in (\ref{uLiederiv04bis}) and (\ref{diff01tot}).) In any case, let us mention that the invariance is restored in the particular case when the macroscopic free spin current $\tau _{\alpha\beta}$ vanishes.

Making use of (\ref{uLiederiv03}), Eq.(\ref{diff01}) reduces to
\begin{eqnarray}
d\,\epsilon ^{\rm u} &=& d\tau\wedge\Bigl[\,{\it l}_u\,\mathfrak{U} - T\,{\it l}_u\mathfrak{s} + E\wedge j + R_{\bot}^{\alpha\beta}\wedge{\tau _{\alpha\beta}}_{\bot}\nonumber\\
&&\hskip1.5cm -\Gamma _{\bot}^{\alpha\beta}\bigl(\,\vartheta _{[\alpha}\wedge\Sigma ^{\rm matt}_{\beta ]}\bigr)_{\bot} \,\Bigr]\,,\label{energycurrder02}
\end{eqnarray}
and finally, comparison of (\ref{energycurrder02}) with (\ref{energycurrder01}), making use of (\ref{spincurrconserv}), yields
\begin{equation}
{\it l}_u\mathfrak{s} -{{\underline{d}\,q}\over T} = {1\over T}\,\bigl[\,E\wedge j + R_{\bot}^{\alpha\beta}\wedge{\tau _{\alpha\beta}}_{\bot} +\Gamma _{\bot}^{\alpha\beta}\bigl(\,D\,\tau _{\alpha\beta}\bigr)_{\bot}\,\bigr]\,.\label{secondlaw}
\end{equation}
In the lhs of (\ref{secondlaw}) we find the rate of change of the entropy 3-form combined in a familiar way with heat flux and temperature. The interpretation of the first term in the rhs is facilitated by the fact that, according to Ohm's law $j=\sigma\,{}^\# E$, it is proportional to $E\wedge j ={1\over\sigma} j\wedge{}^\# j \geq 0$, so that it is responsible for entropy growth. The second term is analogous to the first one. If we suppose that all terms in the rhs of (\ref{secondlaw}) are $\geq 0$, or, in any case, for vanishing macroscopic free spin current $\tau _{\alpha\beta}$, we can consider (\ref{secondlaw}) to be a particular realization of the second law of thermodynamics.

On the other hand, the first law is no other than the conservation equation (\ref{emmattender}) for matter energy, rewritten as (\ref{diff01}) in terms of the internal energy current 3-form (\ref{intenergycurr01}). This reformulation is necessary in order to bring to light the components of $\epsilon ^{\rm u}$ defined in (\ref{intenergycurr02}), that is, heat flux and internal energy respectively, thus making possible to compare the first law with the second one (\ref{secondlaw}) deduced above. (By the way, notice that the inversion of (\ref{intenergycurr01}) to express $\epsilon ^{\rm matt}$ in terms of $\epsilon ^{\rm u}$ suggests to interpret $\epsilon ^{\rm matt}$ as a sort of enthalpy current 3-form.) Making use of (\ref{energycurrder01}), the first law (\ref{diff01}) can be brought to the more compact form
\begin{eqnarray}
{\it l}_u\,\mathfrak{U} -\underline{d}\,q &=& -\bigl(\,{\cal \L\/}_u\,\vartheta ^\alpha\wedge\Sigma ^{\rm matt}_\alpha \bigr)_{\bot} +R_{\bot}^{\alpha\beta}\wedge{\tau _{\alpha\beta}}_{\bot}\nonumber\\
&&+E\wedge j +E\wedge{\it l}_u P +B\wedge{\it l}_u M\,.\label{uLiederiv03bis}
\end{eqnarray}
The first term in the rhs of (\ref{uLiederiv03bis}), that is, the longitudinal part of ${\cal \L\/}_u\,\vartheta ^\alpha\wedge\Sigma ^{\rm matt}_\alpha $, encloses information about mechanic work, whose form depends on the explicit matter energy-momentum  3-form we consider. In particular, by taking it to consist of a pressure term plus an undefined part, as in (\ref{enmom01}), we find
\begin{equation}
\bigl(\,{\cal \L\/}_u\,\vartheta ^\alpha\wedge\Sigma ^{\rm matt}_\alpha\,\bigr)_{\bot} = \bigl(\,{\cal \L\/}_u\,\vartheta ^\alpha\wedge\Sigma ^{\rm undef}_\alpha\,\bigr)_{\bot} +{\cal \L\/}_u\underline{\vartheta}^\alpha\wedge p\,\overline{\eta}_\alpha\,,\label{presscontrib}
\end{equation}
where the last term, in view of (\ref{volLieder}), results to be
\begin{equation}
{\cal \L\/}_u\underline{\vartheta}^\alpha\wedge p\,\overline{\eta}_\alpha = p\,{\it l}_u\overline{\eta}\,,\label{pressderiv}
\end{equation}
being thus identifiable as the ordinary pressure contribution to work as pressure times the derivative of the volume. It is worth remarking that the emergence of this  pressure contribution to the first law does not ocur through derivation of $\mathfrak{U}$ with respect to the volume $\overline{\eta}$ (which is not an independent variable by itself, being defined from the tetrads as (\ref{3deta06})), but with respect to the tetrad components, as in (\ref{Uder02}). Replacing (\ref{presscontrib}) with (\ref{pressderiv}) in the first law equation (\ref{uLiederiv03bis}), we get for it the more explicit formulation
\begin{eqnarray}
{\it l}_u\,\mathfrak{U} -\underline{d}\,q &=& -\bigl(\,{\cal \L\/}_u\,\vartheta ^\alpha\wedge\Sigma ^{\rm undef}_\alpha \bigr)_{\bot}
+R_{\bot}^{\alpha\beta}\wedge{\tau _{\alpha\beta}}_{\bot} +E\wedge j\nonumber\\
&&-p\,{\it l}_u\overline{\eta}+E\wedge{\it l}_u P +B\wedge{\it l}_u M \,,\label{firstlaw01}
\end{eqnarray}
where one recognizes the familiar contributions of internal energy, heat flux and work [including $\bigl(\,{\cal \L\/}_u\,\vartheta ^\alpha\wedge\Sigma ^{\rm undef}_\alpha\,\bigr)_{\bot}$ among the latter ones], together with additional terms. In particular, $E\wedge j$ and the formally similar quantity $R_{\bot}^{\alpha\beta}\wedge{\tau _{\alpha\beta}}_{\bot}$ are present in (\ref{firstlaw01}) due to irreversibility, as read out from (\ref{secondlaw}).

\subsection{General approach}

Let us extend the previous results to the most general scenario in which we modify all matter currents in analogy to $J^{\rm (tot)}$ in order to take into account further possible contributions of a medium. In an attempt to expand the electromagnetic model, we introduce --associated to gravitational interactions-- translational and Lorentz generalizations of the electromagnetic polarization and magnetization of macroscopic matter. Maybe this constitutes a merely formal exercise. However, it can also be understood as a proposal to look for new properties of material media, since we are going to consider the hypothesis of certain new phenomenological matter contributions to the sources of gravity, acting perhaps as dark matter.

Generalizing (\ref{firstcase}), we propose to modify the complete set of field equations (\ref{covfieldeq1})--(\ref{covfieldeq3}) as
\begin{eqnarray}
dH &=&J^{\rm (tot)}\,,\label{covfieldeq1bis} \\
DH_\alpha &=&\Pi ^{\rm (tot)}_\alpha\,,\label{covfieldeq2bis}\\
DH_{\alpha\beta} +\vartheta _{[\alpha }\wedge H_{\beta ]}&=&\tau ^{\rm (tot)}_{\alpha\beta}\,,\label{covfieldeq3bis}
\end{eqnarray}
with bare excitations and total currents consisting of the sum of free and bound contributions, defined respectively as
\begin{eqnarray}
J^{\rm (tot)} &=& J-dH^{\rm matt}\,,\label{Jtot} \\
\Pi ^{\rm (tot)}_\alpha  &=& \Pi _\alpha -DH^{\rm matt}_\alpha \,,\label{Pitot}\\
\tau ^{\rm (tot)}_{\alpha\beta}  &=& \tau _{\alpha\beta} - ( DH^{\rm matt}_{\alpha\beta} +\vartheta _{[\alpha }\wedge H^{\rm matt}_{\beta ]})\,,\label{Tautot}
\end{eqnarray}
where we introduce generalizations of the electromagnetic polarization and magnetization (\ref{matexcit01}) as
\begin{eqnarray}
H^{\rm matt} &=& -d\tau\wedge M + P\,,\label{matexcit01bis}\\
H_\alpha ^{\rm matt} &=& -d\tau\wedge M_\alpha  + P_\alpha \,,\label{matexcit02}\\
H_{\alpha\beta}^{\rm matt} &=& -d\tau\wedge M_{\alpha\beta} + P_{\alpha\beta}\,,\label{matexcit03}
\end{eqnarray}
whatever the physical correspondence of these quantities may be. Since, as discussed above, only matter currents are to be modified, we understand (\ref{Pitot}) in the sense that only the matter part is altered, that is
\begin{equation}
\Pi ^{\rm (tot)}_\alpha = \Sigma ^{\rm matt}_{{\rm (tot)}\alpha } +\Sigma ^{\rm em}_\alpha +E_\alpha\,,\label{totmomentdecomp}
\end{equation} being
\begin{equation}
\Sigma ^{\rm matt}_{{\rm (tot)}\alpha } = \Sigma ^{\rm mat}_\alpha -DH^{\rm matt}_\alpha\,.\label{totmattmom}
\end{equation}
In view of (\ref{totmattmom}), we extend (\ref{mattenergy}) as
\begin{equation}
\epsilon _{\rm (tot)}^{\rm matt} := -u^\alpha\,\Sigma ^{\rm matt}_{{\rm (tot)}\alpha } =\epsilon ^{\rm matt} + u^\alpha DH^{\rm matt}_\alpha\,,\label{totmattenergy}
\end{equation}
and, as a generalization of (\ref{mattender}) to include macroscopic matter, we postulate the formally analogous equation
\begin{equation}
d\,\epsilon _{\rm (tot)}^{\rm matt} = -{\cal \L\/}_u\,\vartheta ^\alpha\wedge\Sigma ^{\rm matt}_{{\rm (tot)}\alpha } -R_{\bot}^{\alpha\beta}\wedge\tau ^{\rm (tot)}_{\alpha\beta} -F_{\bot}\wedge J^{\rm (tot)}\,,\label{genmattender01}
\end{equation}
as the law of conservation of total matter energy. Eq.(\ref{genmattender01}) can be rearranged as
\begin{eqnarray}
&&\mkern-60mu d\,\bigl(\,\epsilon ^{\rm matt} +F\wedge M + T^\alpha\wedge M_\alpha + R^{\alpha\beta}\wedge M_{\alpha\beta}\,\bigr)\nonumber\\
&&= -{\cal \L\/}_u\,\vartheta ^\alpha\wedge\Sigma ^{\rm matt}_\alpha -R_{\bot}^{\alpha\beta}\wedge\tau _{\alpha\beta} -F_{\bot}\wedge J\nonumber\\
&&\quad + d\tau\wedge\Bigl\{ -F_{\bot}\wedge{\it l}_u P +\underline{F}\wedge{\it l}_u M\nonumber\\
&&\hskip1.6cm -T_{\bot}^\alpha\wedge{\cal \L\/}_u P_\alpha +\underline{T}^\alpha\wedge{\cal \L\/}_u M_\alpha\nonumber\\
&&\hskip1.6cm -R_{\bot}^{\alpha\beta}\wedge{\cal \L\/}_u P_{\alpha\beta} +\underline{R}^{\alpha\beta}\wedge{\cal \L\/}_u M_{\alpha\beta}\Bigr\}\,.\nonumber\\
\label{diff03bis}
\end{eqnarray}
(Compare with (\ref{diff02}).) Without going into details, we proceed in analogy to the former case. We define a similar internal energy current 3-form
\begin{equation}
\widehat{\epsilon}^u := \epsilon ^{\rm matt} +F\wedge M + T^\alpha\wedge M_\alpha + R^{\alpha\beta}\wedge M_{\alpha\beta}\,,\label{totintenergy}
\end{equation}
decomposing as
\begin{equation}
\widehat{\epsilon}^{\rm u} =: d\tau\wedge \widehat{q} + \widehat{\mathfrak{U}}\,.\label{intenergycurr02bis}
\end{equation}
Supposing the functional form of $\widehat{\mathfrak{U}}$ to be
\begin{equation}
\widehat{\mathfrak{U}} = \widehat{\mathfrak{U}}\,(\widehat{\mathfrak{s}}\,,P\,,M\,,P_\alpha\,,M_\alpha\,,P_{\alpha\beta}\,,M_{\alpha\beta}\,,\underline{\vartheta}^\alpha \,, u^\alpha\,)\,,\label{uargsbis}
\end{equation}
and with the pertinent definitions analogous to (\ref{Uder01}) and (\ref{Uder02}), first we get
\begin{eqnarray}
{\it l}_u\,\widehat{\mathfrak{U}} &=& \widehat{T}\,{\it l}_u\widehat{\mathfrak{s}} +{\Sigma _{\alpha}}_{\bot}^{\rm matt}\wedge{\it l}_u\underline{\vartheta}^\alpha -\underline{\Sigma}^{\rm matt}_\alpha\,{\it l}_u u^\alpha\nonumber\\
&&-F_{\bot}\wedge{\it l}_u P +\underline{F}\wedge{\it l}_u M\nonumber\\
&&-T_{\bot}^\alpha\wedge{\it l}_u P_\alpha +\underline{T}^\alpha\wedge{\it l}_u M_\alpha\nonumber\\
&&-R_{\bot}^{\alpha\beta}\wedge{\it l}_u P_{\alpha\beta} +\underline{R}^{\alpha\beta}\wedge{\it l}_u M_{\alpha\beta}\,,\label{uLiederiv02bis}
\end{eqnarray}
and finally, suitably rearranging the noncovariant quantities in (\ref{uLiederiv02bis}) into covariant ones defined in analogy to (\ref{thetaLiederiv01}), we arrive at
\begin{eqnarray}
{\it l}_u\,\widehat{\mathfrak{U}} &=& \widehat{T}\,{\it l}_u\widehat{\mathfrak{s}} +{\Sigma _{\alpha}}_{\bot}^{\rm matt}\wedge{\cal \L\/}_u\underline{\vartheta}^\alpha -\underline{\Sigma}^{\rm matt}_\alpha\,{\cal \L\/}_u u^\alpha\nonumber\\
&&-F_{\bot}\wedge{\it l}_u P +\underline{F}\wedge{\it l}_u M\nonumber\\
&&-T_{\bot}^\alpha\wedge{\cal \L\/}_u P_\alpha +\underline{T}^\alpha\wedge{\cal \L\/}_u M_\alpha\nonumber\\
&&-R_{\bot}^{\alpha\beta}\wedge{\cal \L\/}_u P_{\alpha\beta} +\underline{R}^{\alpha\beta}\wedge{\cal \L\/}_u M_{\alpha\beta}\nonumber\\
&&+\Gamma _{\bot}^{\alpha\beta}\Bigl[\,D\,\bigl(\tau ^{\rm (tot)}_{\alpha\beta} -\tau _{\alpha\beta}\bigr) +\vartheta _{[\alpha}\wedge\Sigma ^{\rm matt}_{\beta ]{\rm (tot)}}\Bigr]_{\bot}
\,.\label{uLiederiv04bis}
\end{eqnarray}
Assuming that the analogous of (\ref{spincurrconserv}) holds for generalized matter, that is
\begin{equation}
D\,\tau ^{\rm (tot)}_{\alpha\beta} +\vartheta _{[\alpha}\wedge\Sigma ^{\rm matt}_{\beta ]{\rm (tot)}} =0\,,\label{totspinconserv}
\end{equation}
from (\ref{diff03bis}) with (\ref{totintenergy}) and (\ref{uLiederiv04bis}) follows
\begin{eqnarray}
d\,\widehat{\epsilon}^{\rm u} &=& d\tau\wedge\Bigl[\,{\it l}_u\,\widehat{\mathfrak{U}} -\widehat{T}\,{\it l}_u\widehat{\mathfrak{s}} -F_{\bot}\wedge j + R_{\bot}^{\alpha\beta}\wedge{\tau _{\alpha\beta}}_{\bot}\nonumber\\
&&\hskip1.5cm +\Gamma _{\bot}^{\alpha\beta}\bigl(\,D\,\tau _{\alpha\beta}\bigr)_{\bot}\,\Bigr]\,,\label{diff01tot}
\end{eqnarray}
giving rise, when compared with the differential of (\ref{intenergycurr02bis}), to the second law of thermodynamics with exactly the same form as (\ref{secondlaw}).
Regarding the first law (\ref{diff03bis}) with (\ref{totintenergy})--(\ref{uargsbis}), taking (\ref{enmom01}) as before and using the notation (\ref{Max02}), it takes the form
\begin{eqnarray}
{\it l}_u\,\widehat{\mathfrak{U}} -\underline{d}\,\widehat{q} &=& -\bigl(\,{\cal \L\/}_u\,\vartheta ^\alpha\wedge\Sigma ^{\rm undef}_\alpha \bigr)_{\bot}
+R_{\bot}^{\alpha\beta}\wedge{\tau _{\alpha\beta}}_{\bot} +E\wedge j\nonumber\\
&&-p\,{\it l}_u\overline{\eta}+E\wedge{\it l}_u P +B\wedge{\it l}_u M \nonumber\\
&&-T_{\bot}^\alpha\wedge{\cal \L\/}_u P_\alpha +\underline{T}^\alpha\wedge{\cal \L\/}_u M_\alpha\nonumber\\
&&-R_{\bot}^{\alpha\beta}\wedge{\cal \L\/}_u P_{\alpha\beta} +\underline{R}^{\alpha\beta}\wedge{\cal \L\/}_u M_{\alpha\beta}\,,\label{firstlaw01bis}
\end{eqnarray}
which only differs from (\ref{firstlaw01}) in the additional work contributions corresponding to the gravitational generalizations of polarization and magnetization.

\section{Final remarks}

\subsection{Gravity and conservation of total energy}

Let us examine the role played by gravity in the conservation of energy. In our approach, the first law of thermodynamics can take alternatively the forms (\ref{emmattender}) or (\ref{diff01}), being concerned with the matter energy current either in its form $\epsilon ^{\rm matt}$ or $\epsilon ^{\rm u}$. Differentiation of such matter energy currents generates work expressions, the latter ones acting physically by transforming themselves into different forms of energy. So, mechanic work can produce electric effects, etc. However, these subsequent transformations are not explicitly shown by the thermodynamic equation (\ref{diff01}). Neither the sum of the matter and electromagnetic energy currents is conserved separately, since the addition of (\ref{mattender}) and (\ref{emender}) yields
\begin{eqnarray}
d\,(\epsilon ^{\rm matt}+\epsilon ^{\rm em}) &&= -{\cal \L\/}_u\,\vartheta ^\alpha\wedge (\,\Sigma ^{\rm matt}_\alpha +\Sigma ^{\rm em}_\alpha \,) -R_{\bot}^{\alpha\beta}\wedge\tau _{\alpha\beta}\nonumber\\
&&\neq 0\,.\label{energyconserv02}
\end{eqnarray}
Conservation of energy in an absolute sense, with all possible transformations of different forms of energy into each other taken into account, requires to include also the gravitational energy. Indeed, from (\ref{energyconserv01}) with (\ref{energydec}) we get
\begin{equation}
d\,(\epsilon ^{\rm matt}+\epsilon ^{\rm em}+\epsilon ^{\rm gr})=0\,.\label{energyconserv03}
\end{equation}
This conservation equation, concerned with all forms of energy simultaneously, completes the first law of thermodynamics (\ref{diff01}), which concentrates on the behavior of only the matter energy current $\epsilon ^{\rm u}$. The total energy flux $\epsilon _{\bot}$ in (\ref{energyconserv03}) includes heat flux, Poynting flux in a strict sense and other Poynting-like contributions. The integrated form (\ref{exactform02}) of (\ref{energyconserv03}) can be seen as a sort of generalized Bernouilli's principle.

\subsection{Thermal radiation}

The formalism is not necessarily restricted to gauge theoretically derived forms of energy. It is flexible enough to deal with other thermodynamic approaches, as is the case for thermal radiation, the latter being described not in terms of electromagnetic fields but as a foton gas \cite{Prigogine} \cite{Demirel}. A body in thermal equilibrium is modelized as a cavity filled with a gas of thermal photons in continuous inflow and outflow. The number of photons, the internal energy and the entropy contained in the cavity, the pressure of thermal radiation on the walls and the chemical potential are all functions of the temperature, being respectively given by
\begin{eqnarray}
\mathcal{N} &=& \alpha\,T^3\,\overline{\eta}\,,\label{photgas01}\\
\mathfrak{U} &=& \beta\,T^4\,\overline{\eta}\,,\label{photgas02}\\
T \mathfrak{s} &=& {4\over 3}\,\mathfrak{U}\,,\label{photgas03}\\
p\,\overline{\eta} &=& {1\over 3}\,\mathfrak{U}\,,\label{photgas04}\\
\mu &=& 0\,.\label{photgas05}
\end{eqnarray}
The quantities (\ref{photgas01})--(\ref{photgas05}) automatically satisfy the relation
\begin{equation}
{\it l}_u\,\mathfrak{U} = T\,{\it l}_u\mathfrak{s} -p\,{\it l}_u \overline{\eta}\,,\label{uLiederiv09}
\end{equation}
which constitutes a particular case of the thermodynamic equations found above. Indeed, Eq. (\ref{uLiederiv03}) with vanishing $P$, $M$ and $\tau _{\alpha\beta}$ reduces to
\begin{eqnarray}
{\it l}_u\,\mathfrak{U} = T\,{\it l}_u\mathfrak{s} +{\Sigma _{\alpha}}_{\bot}^{\rm matt}\wedge{\cal \L\/}_u\underline{\vartheta}^\alpha -\underline{\Sigma}^{\rm matt}_\alpha\,{\cal \L\/}_u u^\alpha \,.\label{uLiederiv07}
\end{eqnarray}
By handling the photon gas as matter, and taking for it an energy-momentum  (\ref{enmom03}) with $\widetilde{\Sigma}^{\rm undef}_\alpha =0$ as
\begin{equation}
\Sigma ^{\rm matt}_\alpha = u_\alpha\,\epsilon ^{\rm matt} -d\tau\wedge p\,\overline{\eta}_\alpha\,,\label{enmom04}
\end{equation}
replacement of (\ref{enmom04}) in (\ref{uLiederiv07}) yields
\begin{equation}
{\it l}_u\,\mathfrak{U} = T\,{\it l}_u\mathfrak{s} -p\,{\it l}_u \overline{\eta} +\epsilon ^{\rm matt}_{\bot}\wedge u_\alpha\,T_{\bot}^\alpha\,,\label{uLiederiv08}
\end{equation}
from where, for vanishing torsion, (\ref{uLiederiv09}) follows.

On the other hand, for thermal radiation, the second law (\ref{secondlaw}) reduces \cite{Prigogine} to that of reversible processes
\begin{equation}
{\it l}_u\mathfrak{s} -{{\underline{d}\,q}\over T} = 0\,,\label{revsecondlaw}
\end{equation}
and since the number of photons (\ref{photgas01}) inside the cavity is in general not constant, we propose for this quantity the continuity equation
\begin{equation}
{\it l}_u\mathcal{N} +\underline{d} j_{_N} = \sigma _{_N}\,,\label{photnumber}
\end{equation}
where we introduce $j_{_N}$ as the photon flux and $\sigma _{_N}$ as the rate of photon creation or destruction. Now, from (\ref{photgas01})--(\ref{photgas03}), replacing the values
\begin{equation}
\alpha ={{16\,\pi\,k_B^3\,\zeta (3)}\over{c^3\,h^3}}\,,\qquad \beta ={{8\,\pi ^5\,k_B^4}\over{15\,c^3\,h^3}}\,,\label{alphabeta01}
\end{equation}
with $\zeta $ as the Riemann zeta function, such that $\zeta (3)\approx 1.202$, and being $k_B$ the Boltzmann constant, we get the relation
\begin{equation}
\mathfrak{s} = {4\over 3}\,{\mathfrak{U}\over T} = {{4\beta}\over{3\alpha}}\,\mathcal{N}\approx 3.6\,k_B\,\mathcal{N}\,,\label{alphabeta02}
\end{equation}
so that (\ref{revsecondlaw}) with (\ref{alphabeta02}) yields
\begin{equation}
\underline{d}\,q = T\,{\it l}_u\mathfrak{s} \approx 3.6\,k_B\,T\,{\it l}_u\mathcal{N}\,.\label{photheatflux}
\end{equation}
With (\ref{photnumber}), Eq.(\ref{photheatflux}) transforms into
\begin{equation}
\underline{d}\,q \approx 3.6\,k_B\,T\,(\sigma _{_N} -\underline{d} j_{_N})\,.\label{fluxrelat}
\end{equation}
According to (\ref{fluxrelat}), the divergence of the heat flux $q$ of thermal radiation is proportional to the divergence of the photon flux $j_{_N}$ continuously emitted and absorbed by a body, and it also depends on possible additional contributions $\sigma _{_N}$ due to photon production or destruction.

\section{Conclusions}

We propose an approach to thermodynamics compatible with gauge theories of gravity and beyond. Indeed, the formalism developed in the present paper is explicitly covariant under local Lorentz transformations unless for the symmetry breaking terms present in (\ref{secondlaw}) and (\ref{diff01tot}), (which vanish for $\tau _{\alpha\beta}=0$). Moreover, local translational symmetry as much as local $U(1)$ symmetry are also present in our equations as hidden symmetries, due the particular realization of the Poincar\'e$\otimes U(1)$ gauge group used to derive the field equations and Noether identities which constituted our starting point \cite{Tresguerres:2007ih} \cite{Tresguerres:2002uh} \cite{Tresguerres:2012nu}. In particular, the thermodynamic equations, concerned with the exchange between different forms of energy, are both Poincar\'e and $U(1)$ gauge invariant.

The laws of thermodynamics deduced by us concentrate on the conservation of the matter energy current $\epsilon ^{\rm matt}$ (or, equivalently, $\epsilon ^{\rm u}$), but in addition we complete the scheme giving account of the conservation of total energy, as discussed in Sec. X. In this way we synthesize the total energy balance in classical physics of material media.




\appendix
\section{Eta basis and its foliation}

\subsection{Four-dimensional formulas}

The eta basis consists of the Hodge duals of exterior products of tetrads. One defines
\begin{eqnarray}
\eta &:=&\,^*1 ={1\over{4!}}\,\eta _{\alpha\beta\gamma\delta}\,\vartheta ^\alpha\wedge\vartheta ^\beta\wedge\vartheta ^\gamma\wedge\vartheta ^\delta\,,\label{eta4form}\\
\eta ^\alpha &:=&\,^*\vartheta ^\alpha ={1\over{3!}}\,\eta ^\alpha{}_{\beta\gamma\delta} \,\vartheta ^\beta\wedge\vartheta ^\gamma\wedge\vartheta ^ \delta\,,\label{antisym3form}\\
\eta ^{\alpha\beta}&:=&\,^*(\vartheta ^\alpha\wedge\vartheta ^\beta\,)={1\over{2!}}\,\eta ^{\alpha\beta}{}_{\gamma\delta}\,\vartheta ^\gamma\wedge\vartheta ^\delta\,,\label{antisym2form}\\
\eta ^{\alpha\beta\gamma}&:=&\,^*(\vartheta ^\alpha\wedge\vartheta ^\beta\wedge\vartheta ^\gamma\,)=\,\eta ^{\alpha\beta\gamma}{}_\delta\,\vartheta ^\delta\,,\label{antisym1form}
\end{eqnarray}
with \begin{equation}
\eta ^{\alpha\beta\gamma\delta}:=\,^*(\vartheta ^\alpha\wedge\vartheta ^\beta\wedge\vartheta ^\gamma\wedge\vartheta ^ \delta\,)\,,\label{levicivita}
\end{equation}
as the Levi-Civita antisymmetric object, and where (\ref{eta4form}) is the four-dimensional volume element. With tetrads $\vartheta ^\alpha$ chosen to be a basis of the cotangent space, an arbitrary $p$-form $\alpha$ takes the form
\begin{equation}
\alpha ={1\over{p\,!}}\,\vartheta ^{\alpha _1}\wedge
...\wedge\vartheta ^{\alpha _p}\,(e_{\alpha _p}\rfloor ...
e_{\alpha _1}\rfloor\alpha\,)\,.\label{pform}
\end{equation}
Its Hodge dual is expressed in terms of the eta basis (\ref{eta4form})--(\ref{levicivita}) as
\begin{equation}
\,{}^*\alpha ={1\over{p\,!}}\,\eta ^{\alpha _1 ... \alpha
_p}\,(e_{\alpha _p}\rfloor ... e_{\alpha
_1}\rfloor\alpha\,)\,.\label{dualform}
\end{equation}
Comparison of the variations of (\ref{pform}) with those of (\ref{dualform}) yields the relation
\begin{equation}
\delta \,{}^*\alpha =\,{}^*\delta\alpha -{}^*\left(\delta\vartheta ^\alpha\wedge e_\alpha\rfloor\alpha\,\right) +\delta\vartheta ^\alpha\wedge\left( e_\alpha\rfloor {}^*\alpha\,\right)\,,\label{dualvar}
\end{equation}
analogous to the three-dimensional identities (\ref{formula01}) and (\ref{formula02}) used in the main text.

\subsection{Foliated eta basis}

Let us now make use of (\ref{foliat1}) and (\ref{tetradfoliat}) to calculate
\begin{eqnarray}
\vartheta ^\alpha &=& d\tau\,u^\alpha + \underline{\vartheta}^\alpha\,,\label{teth01}\\
\vartheta ^\alpha\wedge\vartheta ^\beta &=& d\tau\,\Bigl( u^\alpha\,\underline{\vartheta}^\beta - u^\beta\,\underline{\vartheta}^\alpha \Bigr) + \underline{\vartheta}^\alpha\wedge\underline{\vartheta}^\beta\,,\label{teth02}
\end{eqnarray}
etc. Taking then the Hodge duals of (\ref{teth01}), (\ref{teth02}) etc., we find the foliated version of (\ref{eta4form})--(\ref{levicivita}), that is
\begin{eqnarray}
\eta &=& d\tau\wedge\overline{\eta}\,,\label{eta04}\\
\eta ^\alpha &=& -d\tau\wedge\overline{\eta}^\alpha - u^\alpha\,\overline{\eta}\,,\label{eta03}\\
\eta ^{\alpha\beta} &=& d\tau\wedge\overline{\eta}^{\alpha\beta} -\Bigl( u^\alpha\,\overline{\eta}^\beta - u^\beta\,\overline{\eta}^\alpha \Bigr)\,,\label{eta02}\\
\eta ^{\alpha\beta\gamma} &=& -d\tau\,\epsilon ^{\alpha\beta\gamma} -\Bigl( u^\alpha\,\overline{\eta}^{\beta\gamma} + u^\gamma\,\overline{\eta}^{\alpha\beta}
+ u^\beta\,\overline{\eta}^{\gamma\alpha}\Bigr)\,,\nonumber\\
\label{eta01}\\
\eta ^{\alpha\beta\gamma\delta}&=& -\Bigl( u^\alpha\,\epsilon ^{\beta\gamma\delta}-u^\delta\,\epsilon ^{\alpha\beta\gamma}+ u^\gamma\,\epsilon ^{\delta\alpha\beta}- u^\beta\,\epsilon ^{\gamma\delta\alpha}\Bigr)\,,\nonumber\\
\label{eta00}
\end{eqnarray}
where
\begin{eqnarray}
\overline{\eta}&:=& \Bigl( u\rfloor \eta \Bigr) ={1\over{3!}}\,\epsilon _{\alpha\beta\gamma}\, \underline{\vartheta}^\alpha\wedge\underline{\vartheta}^\beta\wedge\underline{\vartheta}^\gamma ={}^{\#}1 \,,\label{3deta06}\\
\overline{\eta}^\alpha &:=&-\Bigl( u\rfloor \eta ^{\alpha}\Bigr) ={1\over{2!}}\,\epsilon ^\alpha{}_{\beta\gamma}\,\underline{\vartheta}^\beta\wedge\underline{\vartheta}^\gamma ={}^{\#}\underline{\vartheta}^\alpha \,,\label{3deta07}\\
\overline{\eta}^{\alpha\beta}&:=&\Bigl( u\rfloor \eta ^{\alpha\beta}\Bigr) =\,\epsilon ^{\alpha\beta}{}_{\gamma}\,\underline{\vartheta}^\gamma ={}^{\#}(\underline{\vartheta}^\alpha\wedge\underline{\vartheta}^\beta\,)\,,\label{3deta08}\\
\epsilon ^{\alpha\beta\gamma}&:=&-\Bigl( u\rfloor \eta ^{\alpha\beta\gamma}\Bigr) =\,u_\mu\,\eta ^{\mu\alpha\beta\gamma}={}^{\#}(\underline{\vartheta}^\alpha\wedge\underline{\vartheta}^\beta\wedge\underline{\vartheta}^\gamma\,)\,,\nonumber\\
\label{3deta09}
\end{eqnarray}
being (\ref{3deta06}) the three-dimensional volume element, such that $\overline{\eta} = u^\alpha\,\eta _\alpha$. Making use of (\ref{thetaLiederiv01})--(\ref{thetaLiederiv04}), (\ref{3deta06}) and (\ref{3deta07}), one can prove that the Lie derivative of this volume can be decomposed as
\begin{eqnarray}
{\it l}_u \overline{\eta} ={\cal \L\/}_u\underline{\vartheta}^\alpha\wedge\overline{\eta}_\alpha\,.\label{volLieder}
\end{eqnarray}
On the other hand, the contractions between tetrads and eta basis in four dimensions (see for instance \cite{Hehl:1995ue}), when foliated reduce to
\begin{eqnarray}
\underline{\vartheta}^\mu\wedge\overline{\eta}_\alpha &=& h^\mu{}_\alpha\,\overline{\eta}\,,\label{rel04bis}\\
\underline{\vartheta}^\mu\wedge\overline{\eta}_{\alpha\beta} &=& -h^\mu{}_\alpha\,\overline{\eta}_\beta +h^\mu{}_\beta\,\overline{\eta}_\alpha\,,\label{rel03bis}\\
\underline{\vartheta}^\mu\,\epsilon _{\alpha\beta\gamma} &=& h^\mu{}_\alpha\,\overline{\eta}_{\beta\gamma} +h^\mu{}_\gamma\,\overline{\eta}_{\alpha\beta} +h^\mu{}_\beta\,\overline{\eta}_{\gamma\alpha}\,,\label{rel02bis}\\
0 &=& -h^\mu{}_\alpha\,\epsilon _{\beta\gamma\delta} + h^\mu{}_\delta\,\epsilon _{\alpha\beta\gamma} -h^\mu{}_\gamma\,\epsilon _{\delta\alpha\beta} + h^\mu{}_\beta\,\epsilon _{\gamma\delta\alpha}\,.\nonumber\\
\label{rel01bis}
\end{eqnarray}
Taking (\ref{dualitycondbis}) into account, we also find
\begin{eqnarray}
e_\alpha\rfloor\overline{\eta}&=&\overline{\eta}_\alpha\,,\label{contract02}\\
e_\alpha\rfloor\overline{\eta}_{\beta}&=&\overline{\eta}_{\beta\alpha}\,,\label{contract03}\\
e_\alpha\rfloor\overline{\eta}_{\beta\gamma}&=&\epsilon _{\beta\gamma\alpha}\,.\label{contract04}
\end{eqnarray}
In view of definition (\ref{3deta09}), the contraction of all objects (\ref{3deta07})-(\ref{3deta09}) with $u_\alpha$ vanishes. From (\ref{3deta07}) then follows that $0=u_\alpha\,\overline{\eta}^\alpha  ={}^{\#}(u_\alpha\,\underline{\vartheta}^\alpha )\,$, thus implying $u_\alpha\,\underline{\vartheta}^\alpha =0\,$.

\section{Plausibility argument}

Let us argue here against the seemingly {\it ad hoc} character of Eqs.(\ref{Uder02}), namely
\begin{equation}
{{\partial\mathfrak{U}}\over{\partial\underline{\vartheta}^\alpha}}={\Sigma _{\alpha}}_{\bot}^{\rm matt}\,,\qquad
{{\partial\mathfrak{U}}\over{\partial u^\alpha}}=-\underline{\Sigma}^{\rm matt}_\alpha\,,\label{condit1bbb}
\end{equation}
showing that, in fact, the internal energy 3-form $\mathfrak{U}$ inherits properties of the original mater Lagrangian, in particular of $L^{\rm matt}_{\bot}$. First we notice that, according to (\ref{intenergycurr02}), $\mathfrak{U}$ is the transversal part of the internal energy current $\epsilon ^{\rm u}$ defined in (\ref{intenergycurr01}) as proportional to $\epsilon ^{\rm matt}$. On the other hand, from the fundamental matter energy-momentum 3-form (\ref{mattenergy}) with (\ref{sigmamatt}) follows
\begin{equation}
\epsilon ^{\rm matt} =\overline{{\cal \L\/}_u\psi}\,\,{{\partial L}\over{\partial d\overline{\psi}}} -{{\partial L}\over{\partial d\psi}}\,\,{\cal \L\/}_u\psi -L^{\rm matt}_{\bot}\,,\label{expmattenergy}
\end{equation}
so that, at least for Dirac matter, we get $\mathfrak{U}= -L^{\rm matt}_{\bot} +$ additional terms.

According to this relation, Eqs.(\ref{condit1bbb}) should resemble the analogous derivatives of $L^{\rm matt}_{\bot}$. In order to calculate them, we make use of the following result proved in \cite{Tresguerres:2007ih}. When considering the foliated Lagrangian density form $L = d\tau\wedge L_{\bot}\,$, depending on the longitudinal and transversal parts of any dynamical variable $Q = d\tau\wedge Q_{\bot} + \underline{Q}\,$, Eq.(D14) of \cite{Tresguerres:2007ih} establishes that
\begin{equation}
{{\partial L}\over{\partial Q}} = (-1)^p\, d\tau\wedge{{\partial L_{\bot}}\over{\partial\underline{Q}}}+{{\partial L_{\bot}}\over{\partial Q_{\bot}}}\,,\label{condit1}
\end{equation}
with $p$ standing for the degree of the $p$-form $Q$. In view of (\ref{condit1}), the matter energy-momentum 3-form defined in (\ref{momentdecompbis}) decomposes as
\begin{eqnarray}
\Sigma ^{\rm matt}_\alpha := {{\partial L^{\rm matt}}\over{\partial \vartheta ^\alpha}} = -d\tau\wedge{{\partial
L_{\bot}^{\rm matt}}\over{\partial\underline{\vartheta}^\alpha}}+{{\partial
L_{\bot}^{\rm matt}}\over{\partial u^\alpha}}\,,\label{condit1b}
\end{eqnarray}
implying
\begin{equation}
{{\partial L_{\bot}^{\rm matt}}\over{\partial\underline{\vartheta}^\alpha}} = -{\Sigma _{\alpha}}_{\bot}^{\rm matt}\,,\qquad
{{\partial L_{\bot}^{\rm matt}}\over{\partial u^\alpha}} = \underline{\Sigma}^{\rm matt}_\alpha\,,\label{condit1bb}
\end{equation}
which reproduce the form of (\ref{condit1bbb}), provided $\mathfrak{U}= -L^{\rm matt}_{\bot}$ as suggested above.

\end{document}